\magnification=1200

\pageno=1
\centerline {\bf INCORPORATING THE SCALE-RELATIVITY PRINCIPLE }
\centerline {\bf IN STRING THEORY AND EXTENDED OBJECTS     }
\medskip
\centerline {\bf Carlos Castro }
\centerline {\bf Center for Particle Theory, Physics Dept.   }
\centerline {\bf University of Texas}
\centerline {\bf Austin , Texas 78712}
\centerline {\bf World Laboratory, Lausanne, Switzerland}
\smallskip

\centerline {\bf November 1996 }
\smallskip
\centerline {\bf ABSTRACT}

First steps in incorporating Nottale's scale-relativity principle to string theory and extended objects are taken. Scale Relativity is to scales what motion Relativity is to velocities. The universal, absolute, impassible, invariant scale under dilatations,  in Nature is taken to be the Planck scale which is not the same as the string scale. Starting with Nambu-Goto actions for strings and other extended objects, we show that the principle of scale-relativity invariance of the world-volume measure associated with the extended objects ( Lorentzian-scalings transformations  with respect to the resolutions of the world-volume coordinates) is compatible  with  the vanishing of the scale-relativity version of the $\beta$ functions : $\beta^G_{\mu\nu}=\beta^X=0$, of the target spacetime metric and coordinates, respectively. Preliminary steps are taken to merge motion relativity with scale relativity and, in this fashion,  analogs of Weyl-Finsler geometries make their appearance. T!
 he quantum case remains to be stud
ied.

\centerline {\bf  I. Introduction}

Despite the tremendous progress in string theory in the past two years a geometrical foundation of the theory, like Einstein's General Theory of Relativity,  is still lacking. Tsetlyn [1] has remarked that the vanishing of the $beta$ functions  for the couplings of the non-linear $\sigma$ model associated with the string propagation in curved backgrounds must contain a clue to the non-perturbative and geometrical 
formulation of string theory.

Some time ago, Amati, Ciafaloni and Veneziano [4] pointed out that some sort of enlarged equivalence principle is operating in string theory in which dynamics is not only independent of coordinate transformations but also of structures ocurring at distances shorter than the string length. In [3] we obtained a modification and extension of the stringy-uncertainty principle within the framework of the theory of special Scale-Relativity proposed by Nottale [2]. In particular, the size of the string was bounded by the minimum Planck scale and an upper, impassible, absolute  scale, invariant under dilations, that incorporates the principle of scale-relativity to cosmology as well. Such principle has allowed Nottale to propose a very elegant and ``simple'' resolution to the cosmological constant problem.

The aim of this letter is far less ambitious. We are able to  incorporate the principle of scale-relativity to the theory of extended objects; i.e. we show that applying this principle to the resolutions of the extended object world-volume coordinates, one can implement Lorentzian-scale relativistic invariance to the 
world-volume measure of these extended objects,  as they are dynamically embedded in a curved target spacetime background. Scale-relativity invariance of areas, volumes,....  
can be maintained if the scale-relativistic version of the $beta$ functions 
$\beta^G_{\mu\nu},\beta^X$ $vanishes$. i.e. scale-relativistic 
transformations preserving  the world-volumes of the extended objects are compatible with the $\beta^G_{\mu\nu}=\beta^{X^\mu} =0$ conditions.   

This is another corroboration, that the theory of extended objects should be interpreted as  a gauge theory of volume preserving diffeomorphisms [11]. It may very well in fact be a composite antisymmetric tensor field theory of the type proposed by 
Guendelman, Nissimov and Pacheva as we have shown in [11]. The relevance of this point of view is that it allowed us to build-in, from the beginning,  the analogs of $S$ and $T$ duality symmetries in extended objects.

After reviewing Nottale's essential results in section {\bf II}.  we 
proceed  into the implementation program of scale-relativity in {\bf II} and {\bf III}. Finally, in the conclusion, we discuss the plausible, and not 
farfetched role,  that Weyl-Finsler geometries might have in achieving the goal of finding the geometrical principle behind string theory and other extended objects. This occurs when we try to merge  motion-relativity with scale-relativity. The quantization program should be carried out to see how far this principle can be taken.

\centerline {\bf II.  }

We shall present  a brief review of Nottale's results. For a detailed account of the theory of Scale Relativity, we recommend the reader to study  Nottale's  work that appeared in [2]. 

Essentially, one has a collection of scalar fields, $\varphi$,  that 
under Lorentzian-scalings (  of the spacetime coordinate $resolutions$  
that our physical apparatus can resolve ) behave like the ordinary spacetime coordinates under ordinary Lorentzian transformations ( change of frame of reference).  The analog of the speed of light, $c$, is played by the logarithm of the $ratio$ of two resolutions : One is the resolution, $\lambda_o$,  with respect to which we measure other resolutions, $\Delta x\leq \lambda_o$; and another is the Planck scale, $\Lambda$ in the appropiate dimension. The analog of time is played by the scaling dimension of the scalar fields, $\varphi$, which Nottale labeled by $\delta$. The origins of scale-relativity were motivated by the fractality of spacetime microphysics and, the very plausible fractality of cosmological structures as well. Ord [2], found that the relativistic quantum mechanics of particles could be reinterpreted in terms of fractal trajectories : continuos but nowhere differentiable in spacetime. 

We do not intend to fall hostage  of the debate of ``what is quantum'' and what is ``classical''. Nottale's view is that a nowhere differentiable spacetime should not be viewed any longer as ``classical''. Our aims are less ambitious. We just go ahead and ask ourselves whether or not one can implement the scale-relativity principle to string theory and extended objects. We start with some definitions :         

The scaling behaviour of $\varphi=\varphi (x,\Delta x)$ under scale-relativistic transformations  is :

$$ln(\varphi /\varphi_o)=ln [(\lambda_0/\Delta x)^{\delta (\Delta x)}]. 
\Rightarrow \varphi (x,\Delta x)=\varphi_o (x)(\lambda_o/\Delta x)^{\delta (\Delta x)}.\eqno (2.1)$$
The scale dimension of the field $\varphi$ in the frame of reference whose resolution is $\Delta x \leq \lambda_o$, where  $\lambda_o$ is a fiducial scale which is $\ge \Lambda$, is :

$$\delta (\Delta x) ={1\over \sqrt {1-{  ln^2 (\lambda_0/\Delta x)\over ln^2(\lambda_0/\Lambda)}}}.~\Delta x \leq \lambda_0. \eqno (2.2)$$
 
Under the Lorentz-scale transformations of the $resolutions$ : $\Delta x\rightarrow \Delta x'$  the  logarithm of $\varphi$ and the scaling dimension $\delta$ transform like the components of a two-vector :

$$ ln(\varphi'/\varphi_o)={ln(\varphi/\varphi_o) -\delta ln \rho \over 
 \sqrt {1-{  ln^2\rho \over ln^2(\lambda_0/\Lambda)}}}.\eqno (2.3)$$

$$\delta'=
   { \delta+  { (ln\rho)( ln {\varphi\over \varphi_0})\over 
ln^2(\lambda_0/\Lambda) }
\over \sqrt {1-{ln^2 \rho \over ln^2(\lambda_0/\Lambda)}}}.
\eqno (2.4 )$$
 
where the composition of dilations ( analog of addition of velocities ) is :

$$ln \rho ={ln(\lambda_o/\Delta x)-ln(\lambda_o/\Delta x')\over 
1-{ln (\lambda_o/\Delta x)(\lambda_o/\Delta x')\over ln^2(\lambda_0/\Lambda)}}.
\eqno (2.5)$$

If one chooses, $\lambda_o =\Delta x $ then in (2.1) $\varphi (x,\Delta x=\lambda_o)=\varphi_o (x)$ the above equations simplify :

$$ln \rho =ln (\Delta x'/\Delta x);~\varphi'(x,\Delta x') =\varphi_o (x) (\Delta x'/\Delta x)^{-\delta'}=
\varphi_o (x) (\lambda_o/\Delta x')^{\delta'}.\eqno (2.6)$$
where :

$$\delta ' =
{\delta (\Delta x) \over \sqrt {1-{ln^2 \rho \over ln^2(\lambda_0/\Lambda)}}}
=\delta (\Delta x) [(1-\beta^2)^{-1/2}].
\eqno (2.7)$$                
with the choice $\delta (\Delta x)=\delta_o (\lambda_o)$ and $\beta \equiv 
[ln(\rho)/ln(\lambda_o/\Lambda)]$ which is the analog of $v/c$ in motion relativity, one can recognize eq-(2.7) as the analog of time dilation in motion-relativity.

We could  set 
$c=ln(\lambda_o/\Lambda)=1$. 
Eq-(2.6) has exactly the $same$ form as (2.1), as it should if covariance is to be maintained. Henceforth, we shall omit the suffix $\Delta x'$. A finite Lorentz-scale transformation implies :

$$ \lambda_o \rightarrow \Delta x.~\varphi_o (x,\lambda_o)\rightarrow \varphi_o (x)e^{-\beta c \delta }.~
\delta_o (\lambda_o)\rightarrow \delta_o (\lambda_o) (1-\beta^2)^{-1/2}.~\beta={ln (\Delta x/\lambda_o)\over ln(\lambda_o /\Lambda)}. \eqno (2.8)$$ 
Then :
$$ln(\varphi/\varphi_o)=-\beta \delta \Rightarrow {\partial ln(\varphi/\varphi_o)\over \partial \delta } =-\beta; 
~{\partial ln(\varphi/\varphi_o)\over \partial \beta } =-\gamma^2\delta=-\gamma^3\delta_o. 
\eqno (2.9)$$
The latter equation shows that $\delta,ln(\varphi/\varphi_o),\beta$ play the same role as time, space coordinates and velocity, respectively,  in motion relativity. Notice the subtlety in the difference upon differentiation w.r.t $\delta$ and $\beta$. 
The analog of a time interval, in a fixed frame of reference where the 
relative ``velocity'' $ln(\rho)=ln(\Delta x'/\Delta x)$ between two-frames remains constant (2.6),  in scale relativity is :

$$\Delta (\delta) \equiv \delta^2 -\delta^1=\delta (\Delta x^2)-
\delta (\Delta x^1);~
\Delta (\delta ') =\delta '^2 -\delta '^1=
\gamma \Delta (\delta);~\gamma =[1-\beta^2]^{-1/2}.\eqno (2.10)$$
and $\beta$ ( relative ``velocity'') is the one given in the r.h.s of (2.7). Extreme caution must be taken in order not
to  confuse the scaling dimension, $\delta$, with its transformation properties under Lorentz-scalings. For example, the quantities $\Delta x,\Delta x'$ can both $flow$ in such a fashion that their $ratios$ remains constant ( imagine scaling both quantities by a common factor) then the quantity    
$\beta =[ln(\rho)/ln(\lambda_o /\Lambda) $ still remains constant .  
This means that the gamma-dilation factor in the r.h.s of (2.7) does not change either. However, the quantity $\delta (\Delta x)$ appearing in the numerator of the expression in the l.h.s of (2.17) does $change$ because $\Delta x$ has flowed.
Both $\Delta x$ and $\Delta x'$ have  flowed in such a fashion that their ratio remained constant. 
A flowing value for $\Delta x$ is not the same as a change of a reference frame One must not confuse the values a coordinate ( in a given frame ) can take with its $transformation$ properties under Lorentz-scale transformations. We shall take $c=1$ from now on and by choosing a frame of reference we mean fixing the value of the relative velocities $v/c=v=\beta=ln (\rho)$ in (2.5,2.6) despite the fact that both quantities $\Delta x, \Delta x'$ can both flow maintaining its ratio fixed. It is in this context that it makes sense to evaluate (2.10).

It is now when it is sensible  to view the scaling dimension as the true analog of a time coordinate. This is not new in string theory. The Liouville mode in non-critical strings has played  the role of a ``temporal'' direction as advocated many times by the authors [10] in connection to the origin of the arrow of time. The quantum phase space origins of a point particle from string solitons and $D$ brane scaffolding dynamics can also be studied within this framework [10].

Similar reasoning applies to the analog of a spatial interval, $\Delta \varphi_=\varphi^2 -\varphi^1$, in the frame of reference where the relative $\beta$ remains fixed, for example. Therefore, in a given frame of reference ( frame for given fixed value of the relative ``velocity'' $\beta$) the scale-relativistic invariant ( the  analog of space-time interval ) is :

$$ d\eta ^2 =[ln^2 (\lambda_0/\Lambda)] (d\delta)^2 -{ (d\varphi)^2 
\over \varphi^2}= (d\delta)^2[1  - {1\over (ln^2 (\lambda_0/\Lambda))}{ (d(ln\varphi))^2 
\over (d\delta)^2 }]=
{(d\delta)^2 \over \gamma^2}\Rightarrow \gamma (d\eta)=d\delta.  \eqno (2.11)$$
For a collection of fields , $\varphi$ all with the same scaling dimensions one has the generalization of flat Minkowski spacetime :
$$ d\eta^2 =[ln^2 (\lambda_0/\Lambda)] (d\delta)^2 -{ \sum_i (d\varphi^i)^2 
\over \sum_i(\varphi^i)2}.\eqno (2.12a)$$
The two-dim metric in (2.11) is flat : $dT^2-dU^2$ with $U=ln(\varphi)$. Similarly, the metric in (2.12a) is also flat as one can see by performing the suitable  change of coordinates  :

$$ { \sum_i (d\varphi^i)^2 
\over \sum_i(\varphi^i)^2}=
\sum_i { (d\varphi^i)^2 
\over \sum_i(\varphi^i)^2}=\sum_i [d(ln \zeta ^i)]^2 =\sum_i (dU^i)^2.$$
$$U^i= ln(\zeta^i) =\int d[ln(\zeta ^i)]= \int {d\varphi^i \over \sqrt { \sum_i (\varphi^i)^2}}.~i=1,2,3....\eqno (2.12b )$$

We are now ready to implement the scale-relativistic transformation to string theory and extended objects; i.e. to the 
Nambu-Goto actions.  Lets take the string case as example. If the action is scale-relativistic invariant

$$ \delta S = {\delta S\over \delta\beta}\delta\beta=0\Rightarrow 
  {\delta S\over \delta\beta}=\int d\sigma^1d\sigma^2 {\delta\over \delta\beta}\sqrt { det~|G_{\mu\nu}\partial_{\sigma_1}X^\mu \partial_{\sigma_2}X^\nu |} =0;\mu,\nu =0,1...D-1.  \eqno (2.13a)$$
with 

$$S=\int d\sigma^1d\sigma^2 \sqrt { det~|G_{\mu\nu}\partial_{\sigma_1}X^\mu \partial_{\sigma_2}X^\nu |}; ~G_{\mu\nu}[(X^\mu(\sigma^1,\sigma^2,\Delta 
\sigma^1,\Delta \sigma^2].  \eqno (2.13b)$$
where the target space time coordinates are chosen now to have an explicit dependence on the 
( logarithm of the ) resolution : $-ln(\lambda_o/\Delta x)=ln(\rho)=\beta c$ . 
At the end of this letter a generalization of the action (2.13b) will be discussed that includes an integration and differentiation  w.r.t the resolutions :$\Delta \sigma^a$. As of now it is more prudent to start with a simpler case.  
One was able to pull the ${d\over d\beta}$ inside the integral (2.13a) because the resolution $\Delta \sigma^a$ in scale-relativity by definition is $independent$ of the world sheet coordinates, $\sigma^1, \sigma^2$. One can now establish the correspondence between the string target-spacetime coordinates ,$X^\mu $, and the scalars $\varphi^i (x,\Delta x)$ and their scaling dimension, 
$\delta (\Delta x)$, as follows : 
$X^\mu \rightarrow \varphi^i (x^1,x^2,\epsilon );\delta (x^1,x^2,\epsilon)$.
 So by choosing $\sigma^1=x^1,\sigma^2 =x^2$ and 
$ln(\Delta x^1/\lambda_o)=ln(\Delta x^2/\lambda_o)=ln(\Delta x/\lambda_o)=\epsilon =\beta c.$ Where $c=ln(\lambda_o/\Lambda)$ ( we'll set $c=1$) .  
One can infer  from eqs- (2.13)  that one has $D-1$ copies of the $\varphi$ field all of which have the $same$  scaling-dimension, $\delta_o(x^1,x^2,\lambda_o)$. The scaling dimension in principle can have a  dependence on the world sheet coordinates, $x^1,x^2$. 
The $X^\mu$ coordinates, the 
$\varphi^i,\delta$ quantities, must obey the string equations of motion 
given in eq-(2.21). $\delta (x,\Delta x)= \gamma (\beta) \delta_o (x,\lambda_o) =\gamma (\beta) \delta_o (\lambda_o)$ is a particular solution when one has a flat Minkowski metric. Adding an extra  $x$ dependence to $\delta_o (\lambda_o)$ does not change the scaling transformation properties w.r.t the resolutions
 given by  eqs-(2.8).

In this framework, by setting the two values of $\Delta x^a$ equal to $\Delta x$, means that the Lorentz-scale transformations of the fields w.r.t the resolution, $\Delta x$, (2.8) are essentially $one$ dimensional. A generalization must include scaling transformations for $\Delta x^1,\Delta x^2,\Delta x^3,....$ that involve the analog of rotations as well as transformations that intertwine the coordinates, $x$,  with the resolutions. Preliminary steps in that direction were taken by Nottale in [2] 
where 
scalings ( induced by the Weyl gauge field ) are linked to translations  :$\delta \tau =
-A_\mu dx^\mu$ and $d\tau^2 =g_{\mu\nu}dx^\mu dx^\nu$. i.e; in Weyl's geometry the lengths are not left invariant but acquire scalings.

A plausible way to generate translations induced by scalings can be achieved by relating the metric to the Weyl potential as follows : Ne'eman and Sijacki [7] were able to induced diffeomorphisms from inter-hadronic QCD interactions by starting with the $SU(3)$ color-neutral two-gluon operator product in $D=4$ :

$$G_{\mu\nu}= (\kappa)^{-2}A^a_\mu A^b_\nu g_{ab}. \eqno (2.14)$$
where $\kappa$ has mass dimensions; $a,b..$ are $SU(3)$ adjoint representation indices and $g_{ab}$ is the Cartan metric for the $SU(3)$ gluon octet : $A^a_\mu$. This procedure allowed Ne'eman and Sijacci to reproduce $effective$ local diffeomorphisms ( which translations are a part of ) :

$$\delta_\zeta G_{\mu\nu}=\partial_\mu \zeta_\nu +
\partial_\nu \zeta_\mu. \eqno (  )$$
where $\zeta_\mu$ is  a field dependent parameter given in terms of the instanton pure-gauge part of the original gluon fields. This mechanism was labeled 
[7] Chromo-Gravity. The relevance of this construction is that it provides  way as to how a local gauge symmetry ( it can be generalized to gauge fields of the Weyl type ) can induce translations.  

After this slight detour, eq-(2.13a)  yields :

$${ d(det~h_{ab})^{1/2} \over d \beta }={1\over 2}(det~h_{ab})^{1/2}h^{ab}
{dh_{ab}\over d \beta} =0 ; ~h_{ab}=G_{\mu \nu}\partial_a X^\mu \partial_b X^\nu. \eqno (2.16) $$

Hence :

$$h^{ab}{dh_{ab}\over d \beta} =h^{ab}\beta^G_{\mu\nu}\partial_a
 X^\mu \partial_b X^\nu +h^{ab}G_{\mu\nu}\partial_a
 \beta^{X^\mu} \partial_b X^\nu +h^{ab}G_{\mu\nu}\partial_a
 X^\mu \partial_b \beta ^{X^\nu} =0. \eqno (2.17)$$
The scale-relativity version of the $beta $ function for the metric is defined as $\beta^G_{\mu\nu}\equiv {dG_{\mu\nu}\over d\beta}$.
The space time coordinates are scalars from the world sheet point of view and their scale-relativity version of the $beta$ function is defined as :

$$   \beta^{X^\mu}(X^\mu,\beta)\equiv 
         -{\partial X^\mu \over \partial ln (\lambda_o/\Delta x)}={\partial X^\mu\over \partial \beta}\sim X^\mu. \eqno (2.18)$$
What appears to be the trivial solutions to eq-(2.17) : $\beta^G_{\mu\nu}=
\beta^X=0$ will turn out to be the $correct$ avenue to pursuit.  Lets imagine that we decide to take a different avenue and look for solutions other than the trivial ones.   
At this point we must emphasize that it is Lorentz-scale invariance/covariance which dictates the form of eq-(2.18). The $X^\mu$ are ``eigenvectors'' of the analog of the dilation operator ${\cal D}=\mu{\partial \over \partial \mu}$.  
For example under scale-relativistic transformations ( dilations)  the $\varphi$ behave as  :

$$\varphi (x,\Delta x) =\varphi_o (x)(\lambda_o/\Delta x)^{\delta (x,\Delta x)}
\Rightarrow \beta^\varphi = {\partial \varphi\over \partial \beta} =
-\gamma^3\delta_o \varphi.
~\beta^\delta ={\partial \delta \over \partial \beta }=\beta\gamma^3 \delta_o.
\eqno (2.19)$$

A particular solution to (2.17) ( other than the trivial one ) could be found by regrouping common factors, and
taking for simplicity, nonvanishing metric-elements only those along the diagonal :

$$h^{ab}[\beta^G_{ii} -2\gamma^3\delta_0 G_{ii}]
\partial_a \varphi^i \partial_b\varphi^i                 
+h^{ab}[\beta^G_{00}+2\beta \gamma^3 G_{00}]
\partial_a (\delta_o)\partial_b (\delta_o)+
$$
$$-h^{ab}G_{ii}[\gamma^3\partial_a (\delta_o)(\varphi^i \partial_b \varphi^i)+a\leftrightarrow b]=0. \eqno (2.20)$$

A solution to the above equation can be found first  by  taking 
$\delta_o(x,\lambda_o)$ to be independent of $x$. This choice is compatible with the equations of motion of the $\varphi^i,\delta$ in the Nambu-Goto action  with respect to the world sheet coordinates in the case that the target spacetime is flat : $G_{\mu\nu}$ is the Minkowski metric. The general equations of motion are [12] :

$$
{\delta S\over \delta X^\mu}=G_{\mu\nu}
[{1\over \sqrt {|det~h_{ab}|}}\partial_a ( \sqrt {|det~h_{ab}|}h^{ab}\partial_b  X^\nu ) 
+\Gamma^\nu_{\sigma\tau}\partial_a X^\sigma \partial_b X^\tau h^{ab} ] =0. \eqno (2.21a)$$
For flat diagonal  metrics , $G_{\mu\nu}$ , like the Minkowski metric $\eta_{\mu\nu}$, the Christoffel symbols $\Gamma$ are zero and the equations for $X^0$ become :

$$  {1\over \sqrt {|det~h_{ab}|}}\partial_a ( \sqrt {|det~h_{ab}|}h^{ab}\partial_b  X^0 )=0. \eqno (2.21b) $$ 
An $X^0$ independent on $\sigma^1,\sigma^2$ is a particular trivial solution of (2.21b). From the world-sheet point of view this is very natural, the scaling dimension of the world-sheet scalars, $\varphi^i$, depends on the resolution and not on the location of the fields. From the target space time view, the situation differs. A constant value of the time coordinate  : $X^0=\delta$,  means that the worldsheet is frozen in time. By constant one means constant w.r.t the location of the points in the worldsheet. As we emphasized earlier, the scaling dimension, $\delta$ in  a fixed frame of reference of constant $\beta$ does flow. Once we include 
 all the variables $\Delta x^a$ into the picture by including  them in the range of integration variables of eq-(2.13b), it won't be too odd in having an $X^0$ independent on the $\sigma^a$ coordinates. In any case, as we said earlier we are exploring an avenue which does not turn out to be the correct one. Nevertheless, we consider it  an important academic exercise   that shall enable us gain clarity at the end.

The solutions of eqs-(2.20,2.21b) determine the motion-dynamical behaviour of the 
$\varphi^i_o (x)$ components as well that of  $\delta_o (x,\lambda_o)$ in terms of the $x$ world-sheet coordinates. 
The scaling-dynamics ( of the resolutions, $\Delta x$) at this stage is dictated by the Lorentz-scale symmetry. 
In this respect, the scaling dynamics is trivial, it is pure gauge. Once motion dynamics is treated on equal footing as scaling dynamics in the General Theory of Scale Relativity, the scaling dynamics will cease to be trivial. 
Despite all this one can infer  from (2.20,2.21) that the value of $\beta^G_{\mu\nu}$ is tied up to the motion-dynamical behaviour of the $\delta_o (x,\lambda_o)$. In this respect, motion-dynamics is already intertwined with scaling-dynamics via the Nambu-Goto actions. 
Hence, choosing for a  particular solution :  $\delta_o (x,\lambda_o) =\delta_o(\lambda_o)$ depending only on the running reference scale, solves eq-(2.21b) for $X^0\equiv \delta (x,\Delta x)=\gamma (\beta)\delta_o (x,\lambda_o)=\gamma \delta_o (\lambda_o)$. For this choice of $\delta_o$ the second line  appearing in (2.21) vanishes as well as the terms multiplying the factor $[\beta^G_{00}+2\beta\delta^2 G_{00}]$
This, in conjunction with  having a diagonal metric, leads straightforwards :

$$h^{ab}[\beta^G_{ii} -2\gamma^3 \delta_o G_{11}]\partial_a\varphi^i \partial_b \varphi^i =0.\eqno (2.22)$$
whose $particular$ solution is :

$$ \beta^G_{ii} -2\gamma^3\delta_o G_{ii}=0 \Rightarrow \beta^G_{ii}=
2\gamma^3\delta_o G_{ii}. \eqno (2.23)$$

The value of $\beta^G_{ii}$ is precisely the correct value for  
the choice of metric in eq- (2.11,2.12) 
$$G_{ii}= -{1\over \sum_i (\varphi^i)^2}\Rightarrow \beta^G_{ii}= 2\gamma^3\delta_o 
G_{ii}. \eqno (2.24)$$
where we have used $\beta^{\varphi^i}=-\gamma^3\delta_o\varphi^i$ in (2.19) to determine (2.24). The converse to eq-(2.24) is not necessarily true.

Thus , after pursuing this avenue,  we have verified in this trivial example ( for the particular choice of solutions to equations of motion, when $\delta_o$ is independent on $x$ 
and  for a flat diagonal metric )  that, upon imposing 
${dS\over d\beta }=0\Rightarrow h^{ab}{ d 
h_{ab}\over d\beta}=0$ , and finding a  $particular$ solution of such condition which is  tantamount to the vanishing condition of the $\beta$ function for the induced world-sheet metric : 
${dh_{ab}\over d\beta}=\beta^h_{ab}=0$, it is possible to    
determine the values of $\beta^G_{ii}$ ( and hence the value of $G_{ii}$). 
This has permitted us  us to check  that the $\varphi^i,\delta$ dependence of the metric 
components, $G_{ii}$, must agree $precisely$ with those of the  original scale-relativistic invariant ( flat)  form  given in eq-(2.12). 
Inotherwords, in this example,  the equation $\beta^h_{ab}=0$ determines  the dependence of the target spacetime metric-components, $G_{ii}$ on the $\varphi$ fields.  
The $G_{00}$ component is constrained by scale-relativistic invariance since the element in (2.12) $(d\eta)^2 =G_{00}(d\delta)^2 +G_{ii}(d\varphi^i)^2$ has to remain invariant under scale-relativistic transformations. This fixes $G_{00}$ to have the value of $1$. The case of curved backgrounds will be studied in 
section {\bf III}. 

Essential in this construction was the use of eqs-(2.19). Inotherwords, the solutions for the metric are constrained by the values of $\beta^\varphi,\beta^\delta$ and these, in turn, are $determined$  by the Lorentz-scaling laws in (2.3,2.4). In the remaining of this letter we shall demonstrate that Lorentz-scale invariance of the area-meausure in (2.16) can be maintained if   $\beta^h_{ab}=0\Rightarrow {dS\over d\beta }=0$. However, this  does not mean that ${dS\over d\beta}=0\Rightarrow \beta^h_{ab}=0$   This is precisely nothing but  a manifestation of the principle of scale-relativistic invariance ( applied to the 
$resolutions$ of the world-sheet coordinates, $\Delta x$ ) of the area 
of  the string . The same argument applies to other extended objects.

What are the drawbacks of this avenue : (i). The metric was chosen to be diagonal and flat ( not surprisingly we ended up with the Minkowski metric. (ii). Choosing $\delta$ independent on $x$ is too restrictive as a solution of (2.21a).
(iii). This method is not going to work for curved backgrounds. (iv). All 
this must be contrasted with what we know from string theory : it is the vanishing, $\beta^G_{\mu\nu}=0$, that determines the values of $G_{\mu\nu}$ for $all$ the metric components. 
From string theory we learnt that the condition of vanishing trace (Weyl)  anomaly : $\beta^G_{\mu\nu}=0$, for the two-dimensional quantum stress energy tensor of the non-linear sigma model, was tantamount to the classical background 
massless fields equations of motion including the dilaton and antisymmetric tensor field. For an extensive list of references see  [1]. 

The metric components given by solutions of Einstein's equations  are given in terms of the target spacetime coordinates , $X^\mu$, and $not$ on the $resolutions$, $\Delta X^\mu$. The $\beta^G_{\mu\nu}=R_{\mu\nu}-{1\over 2}G_{\mu\nu} R-T_{\mu\nu}=0$ used in string theory is the standard Quantum Field Theoretic calculation in non-linear $\sigma$ models associated with  background field methods ( or other methods)  in perturbative $D=2$ QFT. In (2.12b) we saw that the metrics (2.11,2.12) are indeed flat and automatically solve Einstein's equations in the absence of matter.   

In view of all these reasons, we arrive at the realization that the correct choice of solutions to eq-(2.16,2.17) is :

$$\beta^G_{\mu\nu}=\beta^X=0 \Rightarrow
\beta^h_{ab}\equiv {dh_{ab}\over d\beta }=0\Rightarrow {dS\over d\beta}=0. \eqno (2.25)$$

Solutions to $\beta^G_{\mu\nu}=\beta^X=0$  exist which are compatible with the scale-relativistic invariant flat metric given in (2.12); i.e.
the solutions can determine the metric. This is precisely what we are going to achieve in the next section. We shall show that the principle of scale-relativity applied to the resolutions of the world-sheet coordinates can be incorporated into string theory and extended objects. 

Evidence has  been furnished  to support the fact that  the theory of extended objects is a gauge theory of volume preserving diffeomorphisms 
[11]. In particular, we were able to show [11] that all $p$-branes can be seen as composite antisymmetric tensor field theories and the analogs of $S$ and $T$ duality could be built-in from the start.  
If this is true, and 
if scale-relativity is a true geometrical principle operating in these theories, their quantization could be based on having areas, volumes, etc..., as multiples of the Planck area, Planck volume,....[6]. For a review of the arguments leading to the minimum Planck scale see [5].   The fact that higher dimensions than four are essential in the quantum theory of membranes, for example, suggests that it is the $D=11,12,..$ Planck lenghts, $l^{11}_P;l^{12}_P$,  which play the role of minimum distances. Our main concern at the moment has been to show that scale-relativity can be implemented in Nambu-Goto actions. 
Which choice of minimal distance to use should be governed by the number of dimensions of the fundamental theory : $\Lambda =L^D_P$.

\smallskip

\centerline {\bf III } 

\centerline {\bf 3.1 The Callan-Symanzik Equation}

\smallskip

We shall prove that (2.25) is the correct way to solve (2.16,2.17). 
Lets begin  with an action, $I(\beta,\delta, X^i,\partial_a X^\mu,
....; G_{\mu\nu}(X^i,\delta,\beta))$ such that 
under infinitesimal scale-relativistic transformations, $\beta \rightarrow \beta+\delta \beta$  it behaves as :
$$\delta I ={\delta I\over \delta \beta}\delta \beta 
=[-\gamma (\beta,\delta) I]\delta \beta \Rightarrow 
I=I_o e^{-\int \gamma (\beta,\delta) d\beta}.~{dI_o\over d\beta}=0. 
\eqno (3.1)$$
i.e. the scale-relativistic transform of $I_o$ is just the usual exponential-scaling ( like those appearing in (2.8) ) whose exponent is a suitable function of $\beta,\delta$ that must $not$ be confused with the Lorentz-scale dilation factor of eqs-(2.7). 
Such  function can  be determined. Examples of $I_o$ such that 
${dI_o\over d\beta}=0$ $are$ the Nambu-Goto action.  
The analog of the Callan-Symanski equation for a general action $I$  is :

$$ [{\partial \over \partial \beta} +{d G_{ij}\over d \beta}
{\partial \over \partial G_{ij}} +
{dX \over d\beta} {\partial \over \partial  X} 
+\partial_a ({dX \over d\beta}) {\partial \over 
\partial (\partial_a X)}+.... 
+\gamma(\beta,\delta) ] I =({d\over d\beta}+\gamma )I=0. \eqno (3.2)$$
where summation over $0,1,2,....$ indices is implied. The coefficients of the differential operator of (3.2) are just the 
$\beta^G,\beta^X,\partial_a \beta^X,...$. Nottale has also discussed the role of the Callan-Symanzik equation in the context of $beta$ functions  [2]. The stringy input of the scale-relativity principle  was proposed by the author [3].  
An inmediate solution for $I(\beta,G_{ij},\varphi,\delta,\partial \varphi, \partial \delta)$ is :

$$I=I_o e^{-\int \gamma (\beta,\delta )d\beta}\Rightarrow {dI\over d\beta}=
-\gamma (\beta,\delta) I. \eqno (3.3) $$
the ``Callan-Symanzik'' equation (3.2)
is  just but the  rephrasing of the statement that under scale-relativistic transformations the action $I$ transforms as expected : by an exponential scaling. 
If one had chosen the Nambu Goto action, $S$, we would have had 
${dS\over d\beta}=0$. The Nambu Goto action belongs to that class of actions 
whose  exponential weight is zero : $\gamma(\beta,\delta)=0 \Rightarrow I=I_o=S$.     

More general equations than ${dI\over d\beta }\sim I$ are in principle feasible; i.e. with the r.h.s having 
a power series expansion of the form :
$${dI\over d\beta}= \sum_n \alpha_n (\beta,\delta)I^n. \eqno (3.4)$$
At the moment we shall focus on the simplest case.

In conjunction with the scale-relativistic version of the Callan-Symanzik equation one must have the equations of motion ${\delta I\over \delta X^\mu}=0$ that determine the behaviour of $\varphi^i_o (x),\delta_o (x,\lambda_o)$ in terms of $x$,  as well as the $beta$ function equations for the metric. Taking the $ij$ components only, for example  :

$$\beta^G_{ij} (G_{ij}) \equiv {d G_{ij} \over d \beta}
=({\partial \over \partial \beta} +{dX\over d \beta}{\partial \over \partial X})G_{ij}= 
( {\partial \over \partial \beta} + \beta^{X}{\partial \over \partial X})G_{ij}. 
\eqno (3.5)$$
Non-perturbative definitions of  $beta$ functions in ordinary field theories  are not given ( as far as we know). In field theory there are no known definitions of $beta$ functions beyond perturbation theory . The scale-relativity 
version of the $\beta$ function is clearly a defined concept. One  can determine the value of the 
$G_{ij}$ by  solving eq-(3.5). In order to solve the latter one can  make an ansatz of the form :

$$\beta^G_{ij} (\beta,\delta,G_{ij}) =G_{\mu\nu}{\cal M}^\mu_i(\beta,\delta)
{\cal M}^\nu_j (\beta,\delta)+....... \eqno (3.6)$$
where the ellipsis stands for terms with higher order powers in $G_{kl}$. The matrices, ${\cal M}^i_j$,  are functions of $\beta,\delta$ only. These are not arbitrary but are tightly constrained to the behaviour of the $\beta^\varphi,
\beta^\delta$ functions :

$$\beta^i(X^i,\beta,\delta)={dX^i\over d\beta}={\cal A}^i_\mu (\beta,\delta)
X^\mu+......\eqno (3.7)$$
as we saw in (2.9).

Keeping only the first terms in the r.h.s of (3.6) and assuming diagonal matrices only  one can integrate 
eq-(3.5,3.6) :

$$G_{ii}=G^o_{ii}e^{\int {\cal M}^i_i {\cal M}^i_i d\beta}; ~{dG^o_{ii}\over d\beta}=0. \eqno (3.8)$$
$G^o_{ij}$ is a solution of $\beta^G_{ij}=0 $ in (3.5); this is the scale-relativity analog of what is ordinarily required from what we know of string theory.
An example of $G^o_{ij}$ can be borrowed from the flat metric components (2.12b) in the frame of reference where $\beta =0$ : $G^o_{ii}=-{1\over \sum_i (\varphi^i_o)^2}$. The $\varphi^i_o =\varphi^i (x,\Delta x=\lambda_o)$ are the 
$reference$ fields with respect to which we perform a Lorentz-scale transformation. If one chooses the $rest$ frame $\Delta x=\lambda_o$ and the former flows by taking $\Delta x \rightarrow \mu \Delta x$  then the reference scale $\lambda_o$ and $\Lambda$ are rescaled by the same amount so their ratio remains fixed and $c=1$ is maintained and $\beta$ remains zero as well.

Under Lorentz-scale relativistic transformations the metric remains invariant.  
 : 

$$(d\eta)^2=G_{\mu\nu}(X) dX^\mu dX^\nu =G'_{\mu \nu }(X') dX'^\mu dX'^\nu. \Rightarrow $$ 
$$G'_{\mu \nu }(X')= G_{\mu\nu}(X')=G_{\rho\eta}(X){dX^\rho\over dX'^\mu} 
{dX^\eta \over dX'^\nu}. 
\eqno (3.9)$$
where the relationship between the $X'$ and the $X$ 
can be obtained from the Lorentz-scale transformations eqs-(2.3,2.4) :

$$X'^\mu ={\Lambda}^\mu_\nu (\beta,\delta) X^\nu. \eqno (3.10)$$
where the matrices $\Lambda$ play the same role as the Lorentz $SO(D-1,1)$ rotation matrices ones in ordinary 
motion-relativity. 

Hence, performing a Lorentz-scale relativistic transformations to a metric $G^o_{\mu\nu}(X^\mu_o)$, allows to determine  
the matrices ${\cal M}$ in (3.6) directly from eqs-(3.5-3.10)
by simply taking derivatives with respect to $\beta$ and choosing $G^o_{\mu\nu}
[X^\mu_o (x,\delta_o)]$; i.e. $X^i_o =\varphi^i_o (x)$.

$$G^o_{\rho\eta}{\Lambda}^\rho_\mu {d{\Lambda}^\eta_\nu \over d\beta}
+G^o_{\rho\eta}( {d{\Lambda}^\rho_\mu \over d\beta}){\Lambda}^\eta_\nu =
G^o_{\rho\eta } {\Lambda}^\rho_\tau {\Lambda}^\eta_\sigma  {\cal M}^\tau_\mu  
{\cal M}^\sigma_\nu .
\Rightarrow \Lambda {d\Lambda \over d\beta}    
+{d\Lambda \over d\beta}\Lambda ={\Lambda}^2{\cal M}^2.
 \eqno (3.12)$$
The above example shows that the scaling-dynamics
of the metric is trivial and fixed by Lorentz-scale invariance. For example, eq-(2.8) in $D=2$ implies :

$${\partial \varphi\over \partial \varphi_o}=e^{-\beta \delta}\Rightarrow 
G_{11}(\varphi)=G^o_{11}(\varphi_o){d\varphi_o\over d\varphi} 
{d\varphi_o \over d\varphi}= G^o_{11}(\varphi_o) e^{2\beta\delta}
\Rightarrow \beta^G_{11} =2\gamma^3\delta_o G_{11}. 
\eqno (3.13)$$
as expected and the ${\cal M}^1_1$ matrices in this example are fully determined to be $2\gamma^3\delta_o$.
The $\beta^G_{ij}$ is not a scalar under Lorentz-scaling  transformations. This can be seen by simple inspection. $\beta^{G^o_{11}}=0\not=2\delta_o G^o_{11}$ in the rest frame where $\gamma =1$. What is a scale-relativistic invariant
  is the ``Callan-Symanzik equation'' and the area-measure of eq-(2.16). Under a change of frame, $\beta^G,\beta^X,...$ change but in such a way that the combined sum of all the terms in (3.2), and in the r.h.s of (2.16),  remains $invariant$. When other resolutions, $\Delta x^1,\Delta x^2,....$, are independently included and not equal to eachother, $\not=\Delta x$, then one must have the full-fledged theory combining $x$ with $\Delta x$ on equal footing. An action like eq-(3.18) must be the relevant one. No longer one will separate motion dynamics from scaling ones. The action should be invariant under $generalized$ diffs.    
In the same way one that the metric was treated, one  has : 

$$\beta^\mu ={d X^\mu\over d\beta}=X^\nu_o {d \Lambda^\mu_\nu \over d\beta} ={\cal A}^\mu_\rho X^\rho = 
{\cal A}^\mu_\rho X^\nu_o {\Lambda}^\rho_\nu\Rightarrow {\cal A}=
{d\Lambda \over d\beta} {\Lambda}^{-1}. 
\eqno (3.14)$$
since $\delta_o$ and $X_o^i=\varphi ^i _o=\varphi ^i_o  (x)$, defined in the ``rest'' frame where  
$\beta =0$, are independent on $\beta$. Hence, one arrives at the matrix relation given by (3.12,3.14) constraining the ${\cal M}, {\cal A}$ matrices in 
terms  of the analog of the Lorentz rotation matrices $\Lambda$.
In a diagonal basis for the latter matrices, like it happens in (2.3,2.4) when $\varphi =\varphi_o$,  eqs-(3.12,3.14) yield 
$2{\cal A}={\cal M}^2$.   

This means that scale-relativistic invariance/covariance constrains the matrices ${\cal M},{\cal A}$ in such a way  that eqs-(3.1,3.2) have  only the ${\Lambda}$ matrices for independent entries and these are entirely determined ( dictated ) by the Lorentz-scale transformation laws. Therefore, the dilation operator : ${d\over d\beta}$ is completely $fixed$ a priori. The scaling dynamics is then trivial ( pure gauge). 
Plugging  $I=I_o=S$ in the `` Callan-Symanzik'' equation  yields a differential equation of the same type as that in eq- (2.20) which can be solved by the same reasoning. Following it, one arrives at the same result in eq-(2.17). Therefore, what is required first of all is to find solutions to $\beta^G =\beta^X=0$ since these solutions automatically will satisfy (2.16,2.17).

Starting, first, with the rest reference frame where $\beta=0 \Rightarrow $ an arbitrary metric-function  ( a ``constant'' of integration ) $G^o_{\mu\nu}[X^\mu_o]$ is a solution to $\beta^G_{\mu\nu}=0$ and the $X^\mu_o$ are trivial solutions to 
$\beta^X =0$ and, thus, eq-(2.16) is trivially satisfied. If Lorentz-scale invariance is to be maintained,  after a scale-relativistic transformation , eq-(2.16) must remain invariant since it is a scalar w.r.t scale-relativistic transformations. Rigorously speaking one must also include the determinant factor 
appearing in (2.16). This is the scale-relativity version of the vanishing of the trace of the stress energy tensor, $T^a_a =0$. If it vanishes in one frame it ought to vanish in another frame if Lorentz-scale invariance holds. 
Therefore, to generate the whole class of 
solutions for all the metrics and fields that obey eq-(2.16), one simply performs a Lorentz-scale transformation of $G^o_{\mu\nu}$ and $X^\mu_o$ as shown in (3.12,3.14). This however does $not$ tell us what the ``constants'' of integration, $G^o_{\mu\nu},X^\mu_o$ are. We have already mentioned how the latter are obtained. In the frame of reference where $\beta =0$, the 
$G^o_{\mu\nu}[X^\mu_o]$ and $X^\mu_o$ quantities $solely$ depend on the worldsheet coordinates and :

(i). $X^\mu_o$ must obey the field equations ${\delta I\over \delta X^\mu}=0$.
For example, the must obey the eqs-(2.21a) of motion where for $G_{\mu\nu}$ one must set it equal to $G^o_{\mu\nu}$.
(ii) The $G^o_{\mu\nu}$, like the flat Minkowski metric (2.11,2.12a,2.12b) , must obey the ordinary Einstein's equations. 

Having completed the former two-steps, one performs  a Lorentz-scaling  :
$$\delta_o \rightarrow \delta =\gamma \delta_o.~\varphi^i_o \rightarrow 
\varphi^i_o e^{-\beta c \delta}.~G^o_{00}=1\rightarrow G_{00}=1.
$$
$$G^o_{ii} =-{1\over  \sum_i (\varphi^i_o)^2}\rightarrow     
G_{ii} =-{1\over  \sum_i (\varphi^i)^2}. ~(det~h^o_{ab})^{1/2} 
=(det~h_{ab})^{1/2}=invariant. \eqno (3.15)$$ 
Since the end result has been to preserve the measure of integration , $(det~h_{ab})^{1/2}$, under scale-relativistic transformations, one reinforces once again the ideas  [11] that extended objects are gauge-theories of volume preserving diffeomorphisms.     

Hence if   
$\beta^h_{ab}=0 
\Rightarrow {dS\over d\beta }=0.$ The converse is not necessarily true : 
What (2.16) means is that the whole meausure ( volume) 
is preserved and not just the particular components of the world-sheet metric given in terms of the embedding spacetime metric.

So far we have discussed the Nambu-Goto action and shown that is scale-relativistic invariant : it satisfies the ``Callan-Symanzik'' equation for zero weight
if the space of metrics and fields are those obtained from Lorentz-scalings of the fiducial metric and fields : $G^o_{\mu\nu},X^\mu_o$. Other types of actions could be studied and plugged-into eq-(3.1) for suitable weights differing from zero or not . Once the values for the $G_{\mu\nu}$ are determined , from the start, to be of the Lorentz-scale invariant form, and the equations of motion ${\delta I\over \delta X}=0$ are solved that determined the $x$ dependence of the fields, the action $I$ can then  be evaluated explicitly. Eq-(3.1) automatically yields  the values of the weighting  functions by explicit differentiation of the action. And viceversa, for a given expression of the weight, the r.h.s of (3.1) gives the action.  In that respect, eqs-(3.1,3.2) 
are again  trivial  because the Lorentz-scale symmetry
determines, from the beginning,  the scaling dynamics. It is when non-trivial scaling dynamics are introduced that eq-(3.1,3.2 ) cease to be trivial. 
This is characteristic of studying the most general metrics; i.e. curved backgrounds and introducing a  generalized  metric which  has a dependence on $X$ and $\Delta X$.   

\centerline{ \bf 3.2  Curved Backgrounds}
\smallskip

So far the $\varphi^i$ world-sheet scalar fields ( and their common scaling dimension)  were taken as the flat target spacetime coordinates. The most natural way to incorporate ( locally) Lorentz invariance in curved backgrounds is by the introduction of tetrads. If $X$ now represent world coordinates and the 
$\varphi^i$ are tangent space coordinates, the tetrad is defined locally as :

$$E^\mu_i \equiv {\partial X^\mu (\xi) \over \partial \xi^i}. ~
\xi^i (x,\Delta x) \equiv \varphi^i (x,\Delta x).~
{\cal G}_{\mu\nu}[X (\varphi,\delta )]=E^i_\mu E^j_\nu \eta_{ij}; ~E^\mu_i E^j_\mu =\delta_{ij}. \eqno (3.16) $$
Setting $\xi^o (x,\Delta x)=\delta (x,\Delta x)$ allows to compute $E^\mu_o, E^o_\mu$. This is how the metric ${\cal G}_{\mu\nu}$ in curved backgrounds depending on 
$X [\varphi (x,\Delta x),\delta (x,\Delta x)]$ is constructed.

From (3.16) it is straightforward to see by inspection that 

$$\beta^X \equiv {dX \over d\beta}=( \beta^\varphi{\partial  \over \partial \varphi }+ 
\beta^\delta {\partial \over \partial \delta })X \not= \beta^\varphi; \beta^\delta
\eqno (3.17)$$
Because the $\beta^X$ is no longer equal to the old $\beta^\varphi$ , the 
dilation operator of (3.1,3.2,3.5,..) is now $different$, as expected,  and no longer flat metrics, like the $G^o_{\mu\nu}$ in (2.11,2.12) and fields like $X^\mu_o$,  are going to be solutions of $\beta^G_{\mu\nu}=\beta^X=0$. Therefore, the implementation of local Lorentz-scale invariance in curved backgrounds  is achieved once more by finding first the fiducial solutions to 
$\beta^G=\beta^X =0 \Rightarrow \beta^h_{ab}=0$. These are the ones obtained 
by solving Einstein's equations ( that determine a metric) and the equations of motion (2.21) for the $X^\mu$ coordinates  with the spacetime metric obtained from Einstein's equations. The local Lorentz-scaling of the tetrads will leave the ${\cal G}_{\mu\nu}$ invariant while rotating the tangent frame.  
     
Hence, $\beta^X$ appearing in (3.17) has changed, which in turn changes the operator form of $\beta^G$ in (3.5), one will no longer have flat Minkowski metrics ( modulo scale transformations) as fiducial solutions. One will have non-flat solutions as expected. Nevetheless, Nambu-Goto actions  
(2.13b) with non-flat metrics will still be solutions of the Callan-Symanzik 
equation.

So far the scale-relativistic  dynamics has been trivial and completely governed by the Lorentz-scale symmetries. To impart true scaling dynamics one must 
start with the  generalizations of Nambu-Goto actions :    

$${\cal S}=\int d^{p+1}x d^{p+1}(\Delta x) \sqrt 
{ det~|{\cal G}_{MN} \partial_a X^M 
\partial_b X^N |.~                           }
 {\cal G}_{MN}[X^\mu (\varphi (x,\Delta x),\delta (x,\Delta x); \Delta X^\mu]. \eqno (3.18) $$
where the derivatives $\partial_a$  are of the type ${\partial \over \partial x}$ and must include now : 
${\partial \over \partial (\Delta x)}$ . Also the metric has acquired an extra dependence on the $\Delta X [\varphi (x,\Delta x),\delta (x,\Delta x)]$ and the $M,N...$ indices run over the $X$ and $\Delta X$ coordinates.  
The dynamics of the $X's $ w.r.t the $\varphi, \delta$ and, hence, w.r.t the $x$ and the $\Delta x$  is no longer trivial ( fixed by Lorentz-scale symmetry like it was in 
(2.8)). 
It is in this $X,\Delta X$ generalized space that one can proceed to construct the General Theory of Scale Relativity encompassing motion and scale relativity on the same footing and including generalized rotations amongst the 
different components of $\Delta X^\mu, X^\mu$. This applies also to a curved world-sheet with $x^1,x^2,\Delta x^1,\Delta x^2$. We hope that string theory might teach us how to achieve this goal. The action (3.18) is invariant under generelized diffeomorphisms which include the ordinary coordinate diffs and the Lorentz-scalings,  as well as the  mixings of the two types of transformations [7].

A simplification of (3.18) is  to set the generalized metric to be dependent on all the $X$ but a only a few of the $\Delta X$. So, choosing  
$\Delta X= \Delta x$ for the indices 
$0,1,2,.....p+1$ ( an analog of the static gauge in extended objects) 
plus performing  the dimensional reduction  of $\Delta x^1,\Delta x^2,...\Delta x^{p+1} \rightarrow \Delta x$ simplifies matters considerably.  
The generalized metric has acquired , then,  an $explicit$ dependence on $\beta$ in addition to the $X$. $\beta$ defined in (2.8) is  the scale analog of motion-velocity.  These type of metrics are the analogs of Finsler metrics. Metrics which live in the tangent bundle of the manifold and depend on positions as well as velocities ( directions). 

Plugging the ${\cal S}$ given by (3.18) in (3.2), for zero weight,  will yield an equation like (2.20). It  should yield the 
same types of equations :$\beta^{\cal G} =\beta^Y =0$, with $Y=X,\Delta X$,
once the generalized equations of motion ${\delta {\cal S}\over \delta Y}=0$ are included. The main obstacle is that we do not know how to write down the 
Generalization of Einstein's equations that determine the motion-scaling-dynamics of the 
generalized metric, ${\cal G}(X,\Delta X)$ in these generalized $X,\Delta X$ spaces. The knowledge of Weyl-Finsler geometry might give us clues as to how to solve this problem.

Physical applications of Finsler geometries in connection to the $maximal$  proper four-accelerations in string theory have been
discussed by Brandt 
[8] and the role of Conformal Weyl-Finsler structures  has been studied by 
[9]. These Finsler metrics are $no$ mathematical curiosities : these metrics are $imposed$ by Stringy-Physics : ``maximal'' proper four accelerations. Weyl-Finsler Geometry is thus a very natural and plausible 
geometrical setting to start  ( and attempt) to build  the geometrical foundations of string theory. 
Weyl-Finsler geometries allow for the introduction of Torsion as well. Riemannian geometry is recovered in a certain limit. In particular,  Einstein's equations appear in the limit of infinite maximal proper acceleration and Riemannian spacetime. In string theory, the effective action for the massless fields contains arbitary powers of curvature terms with the inclusion of the dilaton and antisymmetric tensor field. An interesting project is to see whether the Torsion terms and the extra corrections to the curvature tensor in Finsler-Weyl geometries can account for the massless fields in string theory besides the graviton .

Our main conclusion is that $\beta^G=\beta^X =0\Rightarrow \beta^h_{ab}=0$ is sufficient to recover the background metric and to check invariance ( of volumes) for  
Nambu-Goto actions of  extended objects under Lorentz-scale transformations. The principle of scale-relativity can be incorporated in string theory and extended objects. A challenging question would be if one can maintained scale-relativity invariance at the $quantum$ level.

\centerline {\bf Acknowledegements}

We are greatfuly indebted to G. Sudarshan, L. Nottale for discussions. This work was possible thanks to an ICSC 
World Laboratory, Lausanne, Switzerland  grant.

\centerline {\bf REFERENCES}

1-. A. Tsetlyn : Int. Jour. Mod. Phys. {\bf A 4} (1989) 1257

2-. L. Nottale : `` Fractal Space Time and Microphysics : Towards the Theory

of Scale Relativity''. World Scientific. 1992

G. Ord : Journ. Physics A. Math General {\bf 16} (1983) 1869.

L. Nottale :Int. Jour, Mod. Phys. {\bf A 4} (1989) 5047

and Int. Jour. Mod. Phys. {\bf A 7} (1992) 4899.

L. Nottale : `` Quantum Mechanics, Diffusion and Chaotic Fractals, vol II. 

Pergamon , 1996.

3-. C. Castro : `` String Theory, Scale Relativity and the Generalized 

Uncertainty Relation `` hep-th/9512044. Submitted to Foundations of Physics 

Letters. 

4-. D. Amati, M. Ciafaloni, G. Veneziano : Phys. Lett {\bf B 197} (1987) 81.

D. Gross, P. Mende : Phys. Letters. {\bf B 197} (1987) 129.

5-. L. Garay : Int. Journ. Mod. Phys {\bf A 10} (1995) 145.

6-. A. Ahtekar, C. Rovelli, L. Smolin : Phys. Rev. Letters {\bf 69} (1992) 237.

7-. Y. Ne'eman, D. Sijacki : Mod. Phys. Lett {\bf A 11} (1996) 217.

8-. H. Brandt : `` Finslerian Space Time `` Contemporary Mathematics Series 

of The American Mathematical Society. {\bf 196} (1996) 273.

9-. T. Aikou, Y. Ichijyo : Rep. Fac. Science, Kagoshima Univ. {\bf 23} 

(1990) 101.

10-. J. Ellis, N. Mavromatos, D. Nanopoulos : Phys. Lett.  {\bf B 293 } (1992)

37. Mod. Phys. Lett {\bf A 10} (1995) 425. 

J. Ellis, N. Mavromatos, D. Nanopoulos : `` $D$ Branes from Liouville Strings `` hep-th/9605046

F. Lizzi, N. Mavromatos : `` Quantum Phase Space From String Solitons ``

hep-th/9611040

11-. E.Bergshoeff, E.Sezgin, Y.Tanni, P.Townsend : Annals of Phys. {\bf 199}

(1990) 340.

E. Guendelman, E. Nissimov, S. Pacheva : `` Volume-Preserving Diffeomorphisms

versus Local Gauge Symmetry''. hep-th/9505128.  

C. Castro : `` p-Branes as Composite Antisymmetric Tensor Field Theories'' 

hep-th/9603117. submitted to Class. Quantum Gravity.

12-M. Duff : Class. Quant. Gravity {\bf 6} (1989) 1577. 

\end